\documentclass[twocolumn,showpacs,preprintnumbers,amsmath,amssymb]{revtex4}

\input psfig.sty

\usepackage{graphicx}
\usepackage{dcolumn}
\usepackage{bm}

\begin{document}

\title{Supernova constraints on decaying vacuum cosmology}

\author{S. Carneiro$^{1,2}$\footnote{Associate Member}} \email{saulo@fis.ufba.br}

\author{C. Pigozzo$^{1}$} \email{cpigozzo@ufba.br}

\author{H. A. Borges$^{1}$} \email{humberto@ufba.br}

\affiliation{$^1$Instituto de F\'{\i}sica, Universidade Federal da
Bahia, 40210-340 Salvador - BA, Brasil\\
$^2$International Centre for Theoretical Physics, Trieste,
Italy}

\author{J. S. Alcaniz$^{3}$} \email{alcaniz@on.br}

\affiliation{$^3$Departamento de Astronomia, Observat\'orio Nacional,
20921-400 Rio de Janeiro - RJ, Brasil}

\date{\today} 

\begin{abstract}
There is mounting observational evidence that the expansion of our
Universe is undergoing a late-time acceleration. Among many
proposals to describe this phenomenon, the cosmological constant
($\Lambda$) seems to be the simplest and the most natural
explanation. However, despite its observational successes, such a
possibility exacerbates the well known $\Lambda$ problem, requiring
a natural explanation for its small, but nonzero, value. In this
paper we consider a cosmological scenario driven by a varying
cosmological term, in which the vacuum energy density decays
linearly with the Hubble parameter, $\Lambda \propto H$. We show
that this $\Lambda(t)$CDM model is indistinguishable from the
standard one ($\Lambda$CDM) in that the early radiation phase is
followed by a long dust-dominated era, and only recently the varying
$\Lambda$ term becomes dominant, accelerating the cosmic expansion.
In order to test the viability of this scenario we have used the
most recent type Ia supernova data, i.e., the High-Z SN Search (HZS)
Team and the Supernova Legacy Survey (SNLS) Collaboration data. In
particular, for the SNLS sample we have found $0.27 \leq
\Omega_{\rm{m}} \leq 0.37$ and $0.68 \leq H_0 \leq 0.72$ (at
2$\sigma$), which is in good agreement with the currently accepted
estimates for these parameters.

\end{abstract}

\pacs{98.80.Es; 95.35.+d; 98.62.Sb}
\maketitle

\section{Introduction}

Over the last years, a considerable number of high-quality
observational data have transformed radically the field of
cosmology. Results from distance measurements of Type Ia supernovae
(SNe Ia) \cite{supernovas,rnew,snls} combined with Cosmic Microwave
Background (CMB) observations \cite{cbr}, dynamical estimates of the
clustered matter \cite{calb}, and age measurements of the oldest
structures \cite{age}, seem to indicate that the simple picture
provided by the standard cold dark matter scenario is not enough.
These observations are usually explained by introducing a new
hypothetical energy component with negative pressure, the so-called
dark energy or quintessence (for recent reviews on this topic, see
\cite{review}). Besides its consequences on fundamental physics, if
confirmed, the existence of this dark component would also provide a
definitive piece of information connecting the inflationary flatness
prediction with astronomical data.

On the other hand, from a purely theoretical viewpoint, the
existence of a dark energy is related to an old problem of quantum
field theories and theoretical cosmology, namely the role of
vacuum in the cosmic evolution \cite{Weinberg}. Arguments of
covariance and symmetry lead to an energy-momentum tensor for the
vacuum of the form $T^{\Lambda}_{\mu\nu}=\Lambda g_{\mu\nu}$,
where $\Lambda$ is a scalar function which, in spatially
homogeneous and isotropic space-times, may be, at most, a function
of time only. Therefore, the vacuum acts as a cosmological term,
that is, as a perfect fluid with negative pressure given (in a
comoving frame) by $p_{\Lambda}=-\rho_{\Lambda}=-\Lambda$ (we work
in units where $M_{P} \equiv (8\pi G)^{-1/2} = c = 1$).

Nevertheless, any tentative estimation of the vacuum energy
density by quantum field theories in flat space-time leads to a
divergent result, and any natural cutoff we may choose to impose
in those calculations leads to a vacuum contribution at least
forty orders of magnitude bigger than the observed limits
\footnote{This difference [$\simeq {\cal{O}}(10^{40})$] appears if
we use a cutoff of the order of the energy scale of the QCD chiral
phase transition, the latest of the cosmological vacuum
transitions. For higher cutoffs, like the Planck mass $M_{P}$, the
difference can be as big as 120 orders of magnitude.}. A possible
way out of this trouble is to postulate some cancellation
mechanism that leads to an exactly null vacuum contribution, the
observed dark energy being due to a genuine cosmological constant
or to other fields like quintessence or any other else.

However, a more careful look at the problem of vacuum energy may suggest
another possibility. The divergent result obtained by quantum field
theories in flat space-time cannot, rigorously speaking, be used in
the context of curved, expanding backgrounds. On the other hand, in
Minkowski space-time the Einstein tensor is null, and, therefore,
any vacuum contribution to the energy-momentum tensor, divergent or
not, should be canceled by a bare cosmological constant in Einstein
equations. Now, if we could obtain the vacuum energy density in the
expanding background, we should subtract the Minkowskian
contribution, obtaining a ``renormalized" vacuum density which would
depend on the curvature, being very high for early times, but
decreasing as the universe expands.

As is well known, the Bianchi identities $G^{\mu}_{\nu;\mu}=0$
lead, via Einstein equations, to $T^{\mu}_{\nu;\mu}=0$, which is
an expression of energy-momentum conservation in the presence of
the gravitational field \footnote{Note that, in principle, the
general covariant structure of the Einstein field equations
($G_{\mu\nu} = T^M_{\mu\nu} + \Lambda g_{\mu\nu}$) is not lost if
one supposes $\Lambda$ to be a scalar function of coordinates
$x^{\mu}$. Actually, if one considers the right-hand side of the
above equations as a conserved, total energy-momentum tensor
$T_{\mu\nu}$, then no changes are needed in the general form of
the Einstein-Hilbert action, although it would be necessary to
know the actual interaction term between the matter and vacuum
sectors.}. In the FLRW space-time, this equation leads to
$\dot{\rho}+3H(\rho+p)=0$, where $\rho$ and $p$ are the total
energy density and pressure, respectively, while $H=\dot{a}/a$ is
the Hubble parameter. By introducing the energy densities and
pressures of vacuum and matter/radiation, one finds
\begin{equation} \label{continuidade}
\dot{\rho}_{\gamma} + 3H(\rho_{\gamma} + p_{\gamma})=-\dot{\Lambda}\;.
\end{equation}
The above equation is equivalent to a continuity equation for matter
in the presence of a source $-\dot{\Lambda}$, meaning that the process of
vacuum decay is concomitant to a process of matter production, a
general feature of the vacuum state of any non-stationary
space-time. We are probably far away from a definite theory of
quantum vacuum in curved backgrounds or from the correspondent
microscopic description of vacuum decay. Alternatively, we can
consider effective, theoretically or empirically motivated, decaying
laws for the vacuum density, exploring its effects by means of
macroscopic equations like (\ref{continuidade}). Such an approach
has an old history in the literature \cite{Bertolami} and a renewed
interest in recent years \cite{lambdat0}.

In this regard, a viable possibility has been proposed, with the
vacuum density decaying as $\Lambda = \sigma H$, where
$\sigma\approx m_{\pi}^3$ has the order of the cube of the energy
scale of the QCD vacuum condensation \cite{Schutz}. By using the
observed values of $m_{\pi}$ and $H$, it is straightforward to
verify that the above law provides a value for $\Lambda$ very close
to the value presently ``observed". Naturally, the theoretical
justification for this decaying law is based on some
phenomenological hypothesis and, as such, needs to be verified from
a fundamental theory viewpoint. However, the important aspect here
is that it leads to cosmological solutions in agreement with the
standard scenario for the evolution of the Universe, as explicitly
shown in \cite{Borges}. The above decaying law, for instance, leads
to an early radiation-dominated phase where the vacuum term and the
photon production are negligible, and where the scale factor and
temperature evolve exactly as in the standard Friedmann solution.
This phase is followed by a matter-dominated decelerating era, long
enough to allow structure formation, and during which the vacuum
density and matter production are dismissable until very recently.
Finally, the Universe switches to an accelerated expansion driven by
the vacuum, which tends asymptotically to de Sitter solution. If we
consider the present relative matter density (the only free
parameter of the model, besides the Hubble constant) around $1/3$,
we also obtain the present age parameter $H_0 t_0\approx1$, which is
in good accordance with current observations \cite{age}.

In this paper we are particularly interested in testing the
viability of the above scenarios in light of the latest supernova
(SNe Ia) data, as provided recently by Riess {\it et al.}
\cite{rnew} and by Astier {\it et al.} \cite{snls}. As is well
known, these two SNe Ia samples constitute the compilation of best
observations made so far and provide the most direct evidence for
the observed late-time acceleration of the universe. We also discuss
other observational quantities, as the deceleration parameter $q$,
the transition redshift $z_{\rm{T}}$ (at which the expansion
switches from a decelerated to an accelerated phase), and the total
expanding age of the Universe. In Sec. II we revise the main
expressions and predictions of the cosmological solution with
$\Lambda=\sigma H$. The observational quantities of the model are
discussed in Sec. III. Sec. IV presents our SNe Ia analysis and a
discussion on the observational constraints. In Sec. V we end the
paper by summarizing our main conclusions.

\section{The model}

In the context of a FLRW space-time with null spatial curvature, the
Einstein equations lead to the Friedmann equation
\begin{equation} \label{Friedmann}
\rho = \rho_{\gamma} + \Lambda = 3 H^2,
\end{equation}
which, together with (\ref{continuidade}), the equation of state for
the matter fields [$p_{\gamma} = (\gamma-1) \rho_{\gamma}$], and a decaying law
for $\Lambda$, completely describes the evolution of the scale
factor and densities. As discussed earlier, we will consider
$\Lambda = \sigma H$, so that by combining the above expressions we
find
\begin{equation} \label{evolucao}
2\dot{H} + 3\gamma H^2 - \sigma \gamma H = 0.
\end{equation}
By imposing the conditions $H>0$ and $\rho_{\gamma}>0$, we can also
obtain the solution for the scale factor \cite{Borges}, i.e.,
\begin{equation} \label{a}
a(t) = C \left[\exp\left(\sigma \gamma t/2\right) -
1\right]^{\frac{2}{3\gamma}},
\end{equation}
where $C$ is the first integration constant and the second one has
been set equal to zero, in order to have $a=0$ at $t=0$. Note that
the vacuum and matter densities are, respectively, given by $\Lambda
= \sigma H$ and $\rho_{\gamma} = \rho - \Lambda = 3H^2 -\sigma H$. Thus, by
using Eq. (\ref{a}), it is also possible to rewrite them as
\begin{equation}\label{rhoa}
\rho_{\gamma} =
\frac{\sigma^2}{3}\left(\frac{C}{a}\right)^{3\gamma/2}\left[1 +
\left(\frac{C}{a}\right)^{3\gamma/2}\right]
\end{equation}
and
\begin{equation}\label{Lambdaa}
\Lambda = \frac{\sigma^2}{3}\left[1 +
\left(\frac{C}{a}\right)^{3\gamma/2}\right].
\end{equation}

\subsection{Radiation-dominated Era}

For the radiation epoch ($\gamma = 4/3$), Eq. (\ref{a}) is given as
\begin{equation}\label{aradiation}
a(t) = C \left[\exp\left(2\sigma t/3\right) - 1\right]^{1/2},
\end{equation}
so that in the limit of small times ($\sigma t\ll 1$), we have
\begin{equation}\label{asmall}
a(t) \approx \sqrt{2C^2\sigma t/3},
\end{equation}
which is the same time dependence as in the standard scenario.
Eqs. (\ref{rhoa})-(\ref{Lambdaa}) reduce now to
\begin{equation}\label{rhoaradiation}
\rho_r = \frac{\sigma^2 C^4}{3a^4} + \frac{\sigma^2 C^2}{3a^2}
\end{equation}
and
\begin{equation}\label{Lambdaaradiation}
\Lambda = \frac{\sigma^2}{3} + \frac{\sigma^2 C^2}{3a^2},
\end{equation}
while, in the limit $a \rightarrow 0$, we have
\begin{equation}\label{rhosmall}
\rho_r = \frac{\sigma^2 C^4}{3a^4} = \frac{3}{4t^2},
\end{equation}
and
\begin{equation}\label{Lambdasmall}
\Lambda = \frac{\sigma^2 C^2}{3a^2} = \frac{\sigma}{2t}.
\end{equation}
From the above expressions, it is straightfoward to see that for
small times the expansion is completely driven by the relativistic
matter with its energy density scaling as $a^{-4}$.

\subsection{Matter-dominated Era}

For the dust phase ($\gamma = 1$), Eq. (\ref{a}) scales as
\begin{equation} \label{adust}
a(t) = C \left[\exp\left(\sigma t/2\right) - 1\right]^{2/3},
\end{equation}
which means that for small times (compared to the present), it can be approximated by
\begin{equation}\label{adustsmall}
a(t) = C(\sigma t/2)^{2/3}.
\end{equation}
Note that the above expression has the same time dependence as in the standard scenario.
This, in other words, amounts to saying that the varying cosmological term starts
dominating only very recently, which guarantees a large enough dust-dominated era.

For this matter-dominated epoch, Eqs. (\ref{rhoa})-(\ref{Lambdaa}) read
\begin{equation}\label{rhodust}
\rho_m = \frac{\sigma^2 C^3}{3a^3} + \frac{\sigma^2
C^{3/2}}{3a^{3/2}}
\end{equation}
and
\begin{equation}\label{Lambdadust}
\Lambda = \frac{\sigma^2}{3} + \frac{\sigma^2 C^{3/2}}{3a^{3/2}}.
\end{equation}
The first term in (\ref{rhodust}) gives the usual scaling of
non-relativistic matter fields, whereas the second term is related
to the production of matter at the expenses of the vacuum decay.
Note that in the limit of large times ($\sigma t \gg 1$), Eq.
(\ref{adust}) leads to the de-Sitter solution, i.e.,
\begin{eqnarray} \label{alarge}
a(t) = C \exp\left(\sigma t/3\right).
\end{eqnarray}
Note also [from (\ref{rhodust}) and (\ref{Lambdadust})] that while  $\rho_m$ tends
to zero $\Lambda$ tends to a genuine cosmological constant.

\begin{figure*}
\centerline{\psfig{figure=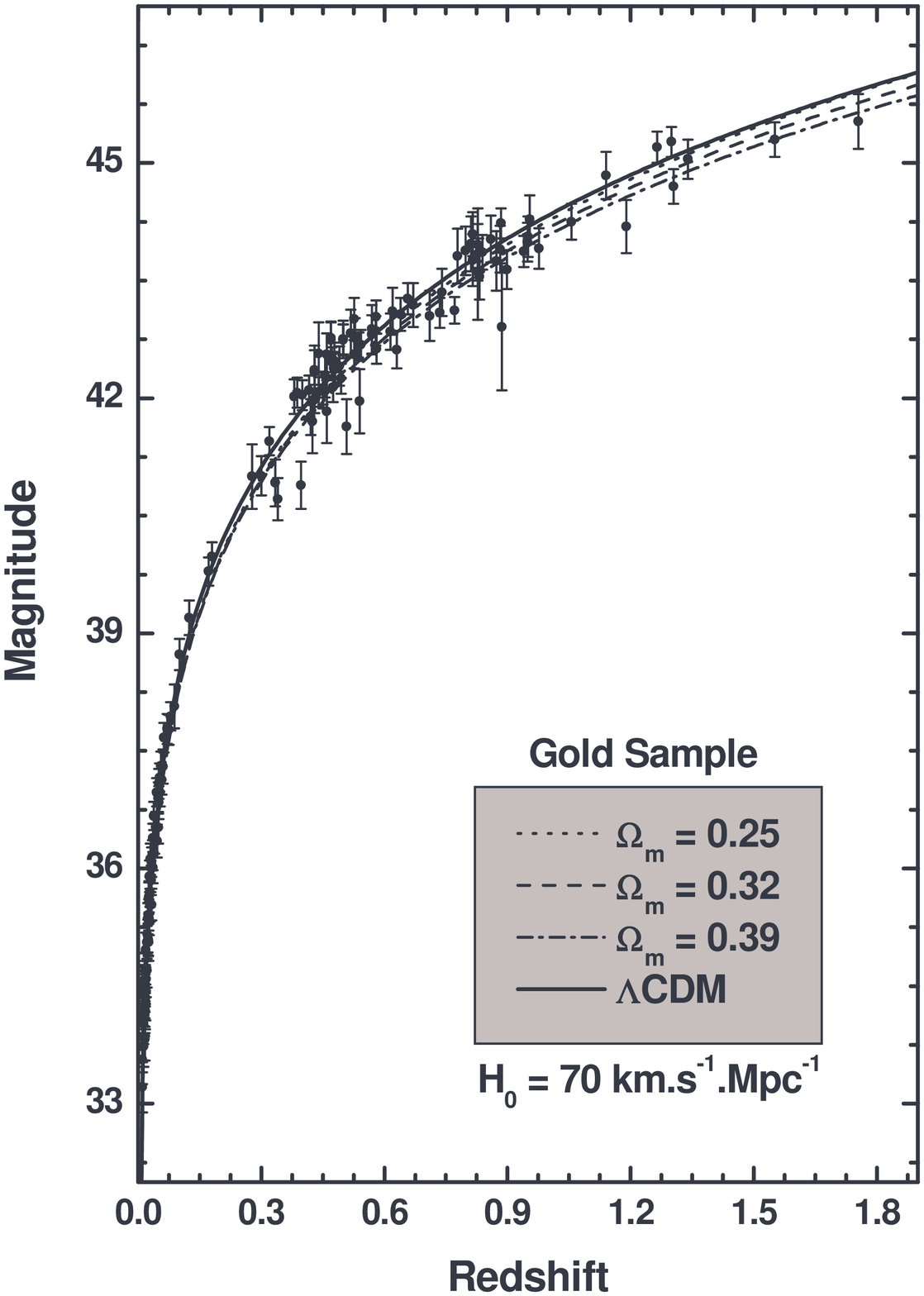,width=3.2truein,height=3.1truein}
\hskip 0.8truecm
\psfig{figure=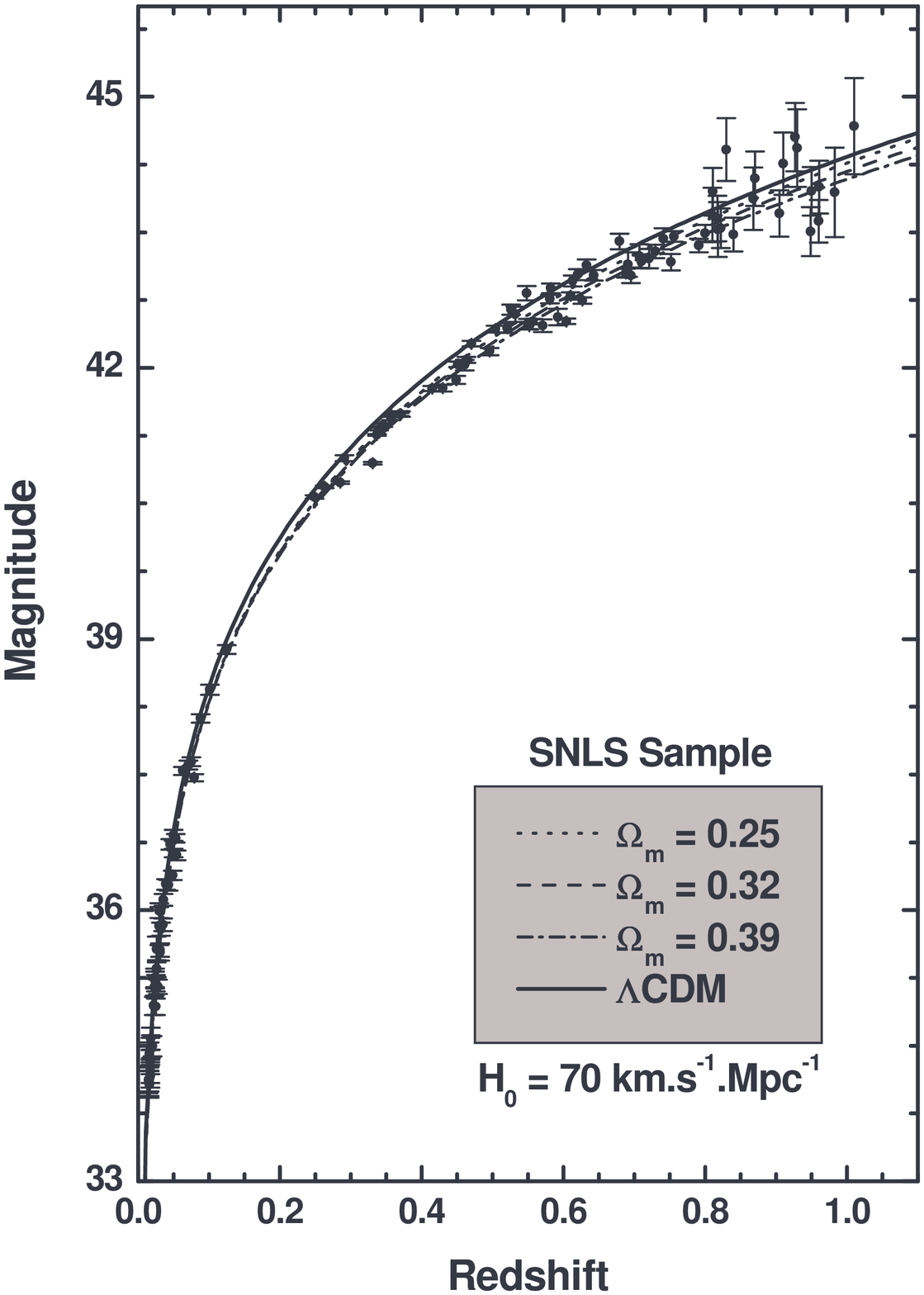,width=3.2truein,height=3.1truein} \hskip
0.1in} \caption{Hubble diagram for 157 supernovae from HZS Team
[Panel (a)] and 115 supernovae from SNLS Collaboration [Panel (b)].
As indicated in the figure, the curves correspond to a fixed value
of the Hubble parameter, $h\, (\equiv H_0/100
\rm{Km.s^{-1}.Mpc^{-1}}) = 0.7$, and selected values of
$\Omega_{\rm{m}}$. For the sake of comparison the current standard
cosmologial model, i.e., a flat $\Lambda$CDM scenario with
$\Omega_{\rm{m}} = 0.27$, is also shown.}
\end{figure*}

\section{Observational quantities}

The Friedmann equation (\ref{Friedmann}) for the dust-dominated
epoch can be rewritten as
\begin{equation} \label{Hz}
H(z) = H_0 \left[1 - \Omega_{\rm{m}} +\Omega_{\rm{m}} (1 + z)^{3/2}\right],
\end{equation}
where $\Omega_{\rm{m}}$ and $H_0$ stand for the current values of
the relative matter density and Hubble parameter. From the above
equation, it is straightforward to show that the deceleration
parameter, defined as $q = -a\ddot{a}/\dot{a}^2$, now takes the
following form
\begin{equation} \label{dec}
q (z)= \frac{\frac{3}{2}\Omega_{\rm{m}}(1 + z)^{\frac{3}{2}}}
{1 - \Omega_{\rm{m}} +\Omega_{\rm{m}} (1 + z)^{3/2}} - 1,
\end{equation}
or, at the present time ($z = 0$),
\begin{equation}
q(z = 0) = \frac{3}{2}\Omega_{\rm{m}} -1.
\end{equation}
Note that for any value of $\Omega_{\rm{m}} < 2/3$ (as indicated by
clustering and dynamical estimates \cite{calb}), the present-day
cosmic expansion is accelerating, which seems to be in full
agreement with current supernova observations \cite{rnew,snls} (see Sec. IV). From
Eq. (\ref{dec}), it is also possible to obtain the transition redshift
$z_{\rm{T}}$ at which the Universe switches from deceleration to
acceleration, i.e.,
\begin{equation} \label{zt}
z_{\rm{T}} = \left[ 2 \left(
\frac{1}{\Omega_m}-1\right)\right]^{2/3}-1.
\end{equation}
As one may anticipate, due to the process of matter production
resulting from the vacuum decay, the transition redshift
$z_{\rm{T}}$ in this model will be always higher than (but of the
same order of) the transition redshift in a $\Lambda$CDM model for
the same value of $\Omega_{\rm{m}}$. In reality, the cosmic
acceleration in the presence of dust matter and a cosmological term
is given by
\begin{equation} \label{acceleration}
6\frac{\ddot{a}}{a}=2\Lambda - \rho_m.
\end{equation}
Therefore, the net effect of the additional terms in (\ref{rhodust})
and (\ref{Lambdadust}) is to increase the acceleration for a given
value of $\rho_m$.

It is also possible to verify that the transition occurs soon after
the first term in (\ref{rhodust}) is supplanted by the sum of the
second term and $\Lambda$. In other words, the late-time
acceleration starts just after the end of the dust matter epoch,
which occurs at the redshift
\begin{equation} \label{z*}
z^* = \left[ (1+\sqrt{2})\left(
\frac{1}{\Omega_m}-1\right)\right]^{2/3}-1.
\end{equation}

Finally, as shown in Ref. \cite{Borges}, the present age parameter can be expressed by
\begin{equation} \label{age}
H_0 t_0 = \frac{\frac{2}{3}\ln(\Omega_{\rm{m}})}{\Omega_{\rm{m}} -
1},
\end{equation}
which means that for the current accepted interval for the matter
density parameter $\Omega_{\rm{m}} = 0.30 \pm 0.05$ ($2\sigma$)
\cite{calb}, one finds $H_0 t_0 \simeq 1.15 \pm 0.08$, i.e., in
accordance with current age parameter estimates \cite{age}.

\section{Supernova Contraints}

SNe Ia observations are certainly among the most remarkable findings
of modern observational cosmology and provide the most direct
evidence for the observed late-time cosmic acceleration. In this
Section we test the viability of the decaying vacuum scenario
discussed above through a statistical analysis involving the most
recent SNe Ia data, namely, the High-Z SN Search (HZS) Team
\cite{rnew} and the Supernova Legacy Survey (SNLS) Collaboration
data \cite{snls}.

\subsection{SNe Ia Samples}

The total sample presented by the HZS Team  consists of 186 events
distributed over the redshift interval $0.01 \lesssim z \lesssim
1.7$ and constitutes the compilation of the best observations made
so far by them and by the Supernova Cosmology Project plus 16 new
events observed by Hubble Space Telescope (HST). This total
data-set was divided into ``high-confidence'' (\emph{gold}) and
``likely but not certain'' (\emph{silver}) subsets.  Here, we will
consider only the 157 events that constitute the so-called
\emph{gold} sample.

The current data from SNLS collaboration correspond to  the first
year results of its planned five years survey. The sample includes
71 high-$z$ SNe Ia in the redshift range $0.2 \lesssim z \lesssim 1$
and 44 low-$z$ SNe Ia compiled from the literature but analysed in
the same manner as the high-$z$ sample. This data-set is arguably
(due to multi-band, rolling search technique and careful
calibration) the best high-$z$ SNe Ia compilation to date, as
indicated by the very tight scatter around the best fit in the
Hubble diagram and a careful estimate of systematic uncertainties.
Another important aspect to be emphasized on the SNLS data is that
they seem to be in a better agreement with WMAP results than the
\emph{gold} sample (see, e.g., \cite{paddy} for a discussion). In
what follows we briefly outline the main assumptions for our
analysis (see also \cite{sne} for some recent SNe Ia analysis).

\begin{figure*}
\vspace{.2in}
\centerline{\psfig{figure=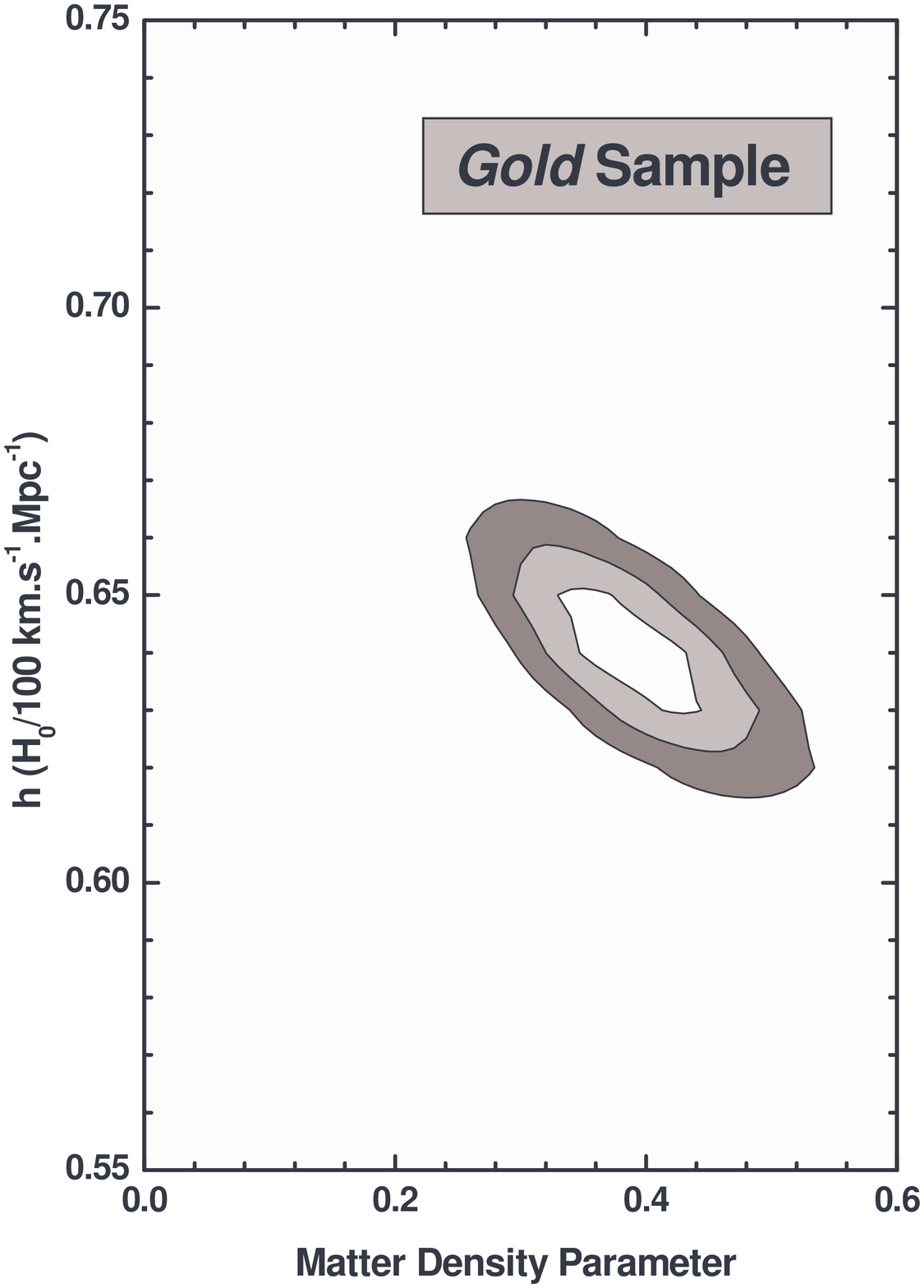,width=3.2truein,height=3.1truein}
\psfig{figure=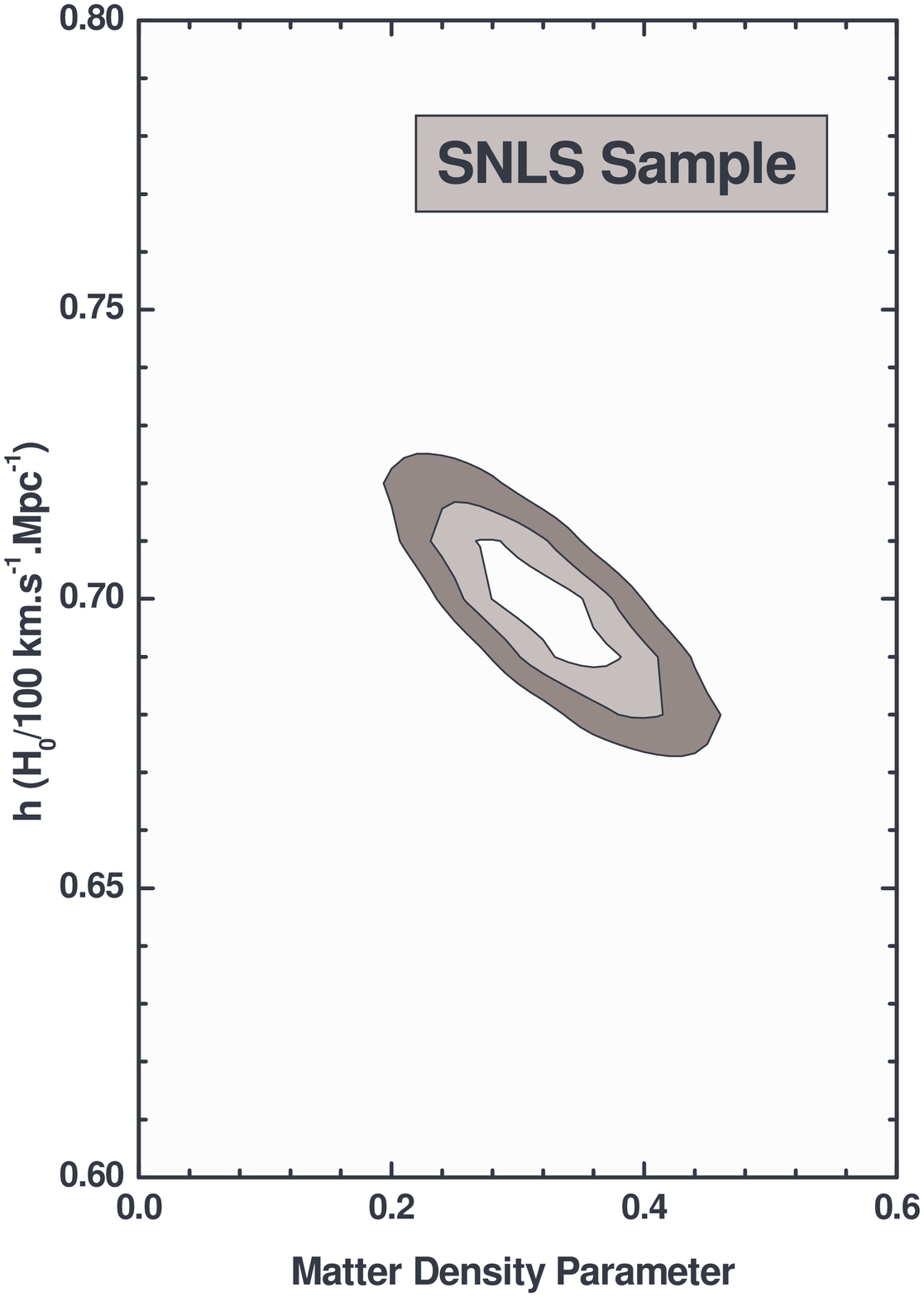,width=3.2truein,height=3.1truein} \hskip
0.1in} \caption{The results of our statistical analyses. ({\bf{Panel
a}}) Confidence regions in the $\Omega_{\rm{m}} - h$ plane for the
\emph{gold} sample of 157 SNe Ia. The best-fit parameters for this
analysis correspond to $\Omega_{\rm{m}} = 0.39$ and $h = 0.64$.
({\bf{Panel b}}) The same as in Panel (a) for the SNLS sample of 115
SNe Ia. In this case, the best-fit model happens at $\Omega_{\rm{m}}
= 0.32$ and $h = 0.70$, with $\chi^{2}_r \simeq 1.0$.}
\end{figure*}

\subsection{Statistical Analysis}

The predicted distance modulus for a supernova at redshift $z$, given a set of
parameters $\mathbf{s}$, is
\begin{equation} \label{dm}
\mu_p(z|\mathbf{s}) = m - M = 5\,\mbox{log}\, d_L + 25,
\end{equation}
where $m$ and $M$ are, respectively, the apparent and absolute
magnitudes, the complete set of parameters is $\mathbf{s} \equiv
(H_0, \Omega_{m})$, and $d_L$ stands for the luminosity distance (in
units of megaparsecs),
\begin{equation}
d_L = c(1 + z)\int_{x'}^{1} {dx
\over x^{2}{H}(x;\mathbf{s})},
\end{equation}
with ${H}(x; \mathbf{s})$ being the expression given by Eq. (\ref{Hz}).

We estimated the best fit to the set of parameters $\mathbf{s}$ by using a $\chi^{2}$
statistics, with
\begin{equation}
\chi^{2} = \sum_{i=1}^{N}{\frac{\left[\mu_p^{i}(z|\mathbf{s}) -
\mu_o^{i}(z)\right]^{2}}{\sigma_i^{2}}},
\end{equation}
where $N = 157$ and $115$ for \emph{gold} and SNLS samples,
respectively, $\mu_p^{i}(z|\mathbf{s})$ is given by Eq. (\ref{dm}),
$\mu_o^{i}(z)$ is the extinction corrected distance modulus for a
given SNe Ia at $z_i$, and $\sigma_i$ is the uncertainty in the
individual distance moduli.

Figures (1a) and (1b) display the Hubble diagram for a fixed value
of the Hubble parameter, $h\, (\equiv H_0/100
\rm{Km.s^{-1}.Mpc^{-1}}) = 0.7$, and some selected values of
$\Omega_{\rm{m}}$. For the sake of comparison our current standard
model, i.e., a flat $\Lambda$CDM scenario with $\Omega_{\rm{m}} =
0.27$, is also shown. Note that for the interval of
$\Omega_{\rm{m}}$ considered the predicted magnitude-redshift
relation is very similar in both classes of scenarios, i.e.,
$\Lambda$CDM and $\Lambda(t)$CDM. In Figs. 2a and 2b we show the
results of our statistical analysis. Confidence regions ($68.3\%$,
$95.4\%$ and $99.7\%$) are shown in the $\Omega_{\rm{m}} - h$ plane
by considering the \emph{gold} and SNLS samples respectively. The
best-fit parameters for the \emph{gold} analysis are
$\Omega_{\rm{m}} = 0.39$ and $h = 0.64$, with the reduced
$\chi^{2}_r \equiv \chi_{min}^2/\nu \simeq 1.15$ ($\nu$ is defined
as degrees of freedom). At 95\% c.l. we also obtain the intervals
$0.34 \leq \Omega_{\rm{m}} \leq 0.44$ and $0.62 \leq h \leq 0.66$.
Note that these estimated values for the matter density parameter
are considerably large, being only marginally compatible with the
current accepted range, i.e., $\Omega_{\rm{m}} = 0.30 \pm 0.05$
($2\sigma$) \cite{calb}. We also note that the above reduced value
of $\chi^2$ is very similar to the one found for the flat
$\Lambda$CDM scenario and is slightly larger than the one obtained
for the $\Lambda$CDM model with arbitrary curvature ($\chi^{2}_r
\simeq 1.12$). For the SNLS analysis (Fig. 2b), the best-fit
parameters are $\Omega_{\rm{m}} = 0.32$ and $h = 0.7$ ($\chi^{2}_r
\simeq 1.0$), which correspond to an accelerating universe with $q_0
\simeq -0.52$, a total expanding age $t_0 \simeq 15.7$ Gyr, and a
transition redshift $z_{\rm{T}} \simeq 1.62$ (with $z^* \simeq
1.97$). At 95\% c.l. we also obtain $0.27 \leq \Omega_{\rm{m}} \leq
0.37$ and $0.68 \leq h \leq 0.72$, which seem to be in a better
agreement with the current accepted values for both clustered matter
density and Hubble parameters.

\section{Conclusions}

We have discussed a decaying vacuum scenario which is
indistinguishable from the standard model with a genuine
cosmological term in what concerns the general features of the
predicted cosmic evolution. The early radiation phase in this
$\Lambda(t)$CDM model is unaffected by the process of vacuum decay,
as well as the physical phenomena taking place at early times (e.g.,
the primordial nucleosynthesis). The following era is dominated by
dust for a long time, and only recently the varying
cosmological term has become important.

Here, we have presented some quantitative results which clearly show
that, even in the current stage of the Universe evolution, our
decaying vacuum scenario is very similar to the standard one. We
have also statistically tested the viability of the model by using
the most recent SNe Ia observations, as given in Refs.
\cite{rnew,snls}. For the so-called \emph{gold} sample, we have
found $0.34 \leq \Omega_{\rm{m}} \leq 0.44$ and $0.62 \leq H_0 \leq
0.66$ at 95.4\% c.l., with the reduced $\chi^2_{r} \simeq 1.15$,
which is similar to the one obtained in Ref. \cite{rnew} for a flat
$\Lambda$CDM model. The SNLS data in turn provide $0.27 \leq
\Omega_{\rm{m}} \leq 0.37$ and $0.68 \leq H_0 \leq 0.72$ (with
$\chi^{2}_r \simeq 1.0$), which is in better agreement with the
currently accepted estimates for both parameters. From the analysis
presented above, we also noted that a more precise determination of
the transition redshift $z_T$ from upcoming SNe Ia data may be able
to distinguish this scenario from the standard model since their
predictions for this quantity are considerably different [see Eqs.
(\ref{zt}) and (\ref{acceleration})].

We also emphasize that an important observational aspect that
deserves a careful investigation concerns the growth of density
perturbations in the realm of this $\Lambda$(t)CDM model. In this
regard, a preliminary analysis indicates that the evolution of the
matter contrast $\delta\rho/\rho$ shows no considerable difference
relative to the standard $\Lambda$CDM case. A complete study on
the formation of large-scale structures in this class of scenarios
will appear in a forthcoming communication.

Finally, it is worth observing that, from the theoretical
viewpoint, it would be interesting to investigate possible
relations between our approach and other quintessence models or
modified gravitational theories recently discussed in the
literature. For this purpose, we note that some authors have
already pointed out a mathematical equivalence between dark energy
and scalar-tensor or other forms of ideal fluid with inhomogeneous
equations of state (see, e.g., \cite{arbitro}).

\section*{Acknowledgements}

The authors are very grateful to J. A. S. Lima for useful discussions.
CP and HAB are supported by Capes. JSA is supported by CNPq
(Brazilian Agency), under Grants No. 307860/2004-3 and
475835/2004-2, and by Funda\c{c}\~ao de Amparo \`a Pesquisa do
Estado do Rio de Janeiro (FAPERJ), No. E-26/171.251/2004.

\end{document}